**Title: Oscillating currents stabilize aluminium cells for efficient, low carbon production**

**Author List:** Ibrahim Mohammad(1), Marc Dupuis(2), Paul D. Funkenbusch(1), Douglas H. Kelley(1)

1- Department of Mechanical Engineering, University of Rochester, Rochester, NY 14627
2- GeniSim Inc., Jonquière, Québec, Canada, G7S 2M9

**Abstract**

Humankind produced 63.7 million metric tons of aluminium in 2019[1], nearly all via an electrochemical process in which electrical current liberates molten Al from dissolved alumina. That year, Al production required 848 TWh of electricity[1], 3% of the worldwide total[2], and caused 1% of human greenhouse gas emissions[3]. Much of the electricity and emissions originate from energy loss in the poorly conducting electrolyte where aluminum oxide is dissolved. Thinning the electrolyte layer could decrease loss[4] but has been limited by the Metal Pad Instability (MPI)[5], which causes Al cells to slosh out of control if the electrolyte is not sufficiently thick. Here we show that adding an oscillating component to the current disrupts the MPI in realistic simulations, allowing stable operation with electrolyte layers at least 12% thinner. This occurs when oscillation excites standing waves, which decouple the resonance that drives a growing traveling wave, characteristic of the MPI. Maintaining oscillation can prevent MPI; initiating oscillation can halt an MPI in progress. Our findings could significantly increase the efficiency of virtually all aluminium refining cells without the need for expensive reconstruction, thereby decreasing energy use by 34 TWh/year (2.1 MJ/kg Al) or more and greenhouse gas emissions by 13 Mton/year or more.

**Main**

Aluminium is produced using Hall-Héroult electrolysis cells that contain two broad (~8 x 3.6 m, or larger) and shallow fluid layers (~5-20 cm): molten Al beneath a floating layer of molten cryolite electrolyte, in which raw aluminium oxide is dissolved. A large, steady current (~$10^5$ A) is driven downward from a carbon anode at the electrolyte surface to a cathode beneath the aluminium (Fig. 1a), reducing the aluminium oxide to Al and producing carbon dioxide ($CO_2$), an important greenhouse gas, at the anode. Substantially more $CO_2$ is emitted when generating the necessary electricity, most of which comes from fossil fuels[6]. Of the 51 Gton/year of $CO_2$–equivalent ($CO_2$e) emissions produced by humankind[7], 1% comes from Al production[3], and 62% of that from the electricity[6]. To meet growing global demand while decreasing the $CO_2$ emissions that exacerbate climate change, it is critical to increase the energy efficiency of Al production.

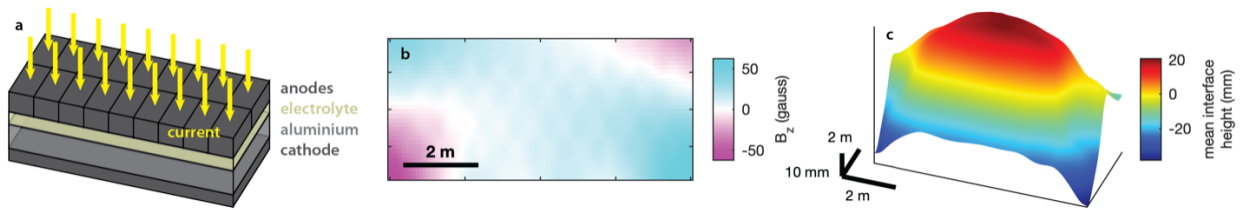

**Figure 1**: Characteristics of Al electrolysis cells. **a,** Electrical current flows downward from anodes to cathode, through electrolyte and Al layers. **b,** The vertical component of the ambient magnetic field, at the Al-electrolyte interface, from our simulations, seen from above. The field varies spatially and is caused primarily by currents in the cell and nearby busbars. **c,** The Al-electrolyte interface bulges because of electromagnetic forces due to the field and the current.

About 40% of a cell's electrical energy reduces no Al[4], instead being transformed into heat by resistance in the electrolyte, which is four orders of magnitude less conductive than Al. Being approximately proportional to electrolyte layer thickness, this loss can be decreased by thinning the layer. However, when the thickness is decreased below a certain critical threshold, the cell becomes unstable[8-12]. In a process known as the metal pad instability (MPI), electromagnetic forces amplify small perturbations on the Al-electrolyte interface, causing a circulating traveling wave that can grow exponentially until the cell sloshes out of control or the Al shorts to the anode. To predict the threshold, multiple theories have been proposed [8-10,13,14], depending in part on the uniformity of the ambient vertical magnetic field (Fig. 1b). In all theories, fluid instability imposes a minimum electrolyte thickness. Many methods for MPI suppression have been attempted in the past, including inserting baffles in the Al layer[4], or tilting the anode in synchrony with interface motion[4,15], with limited success.

In practice, the MPI is mitigated by keeping the electrolyte layer thick and building cells with much greater length $L_x$ than width $L_y$. A large aspect ratio hinders the MPI because it is a parametric instability that depends on a coupled resonance between interface wave modes[17]. The height of the Al-electrolyte interface varies spatially and can be approximated as

$$\sum_{m,n} \alpha_{m,n} G_{m,n} = \sum_{m,n} \alpha_{m,n} \cos\frac{m\pi}{L_x}\left(x + \frac{L_x}{2}\right) \cos\frac{n\pi}{L_y}\left(y + \frac{L_y}{2}\right), \tag{1}$$

where $G_{m,n}$ is an interface mode, $\alpha_{m,n}$ is its amplitude, $(m, n)$ are non-negative integers, $x$ increases along the long axis of the rectangular cell, $y$ increases along its short axis, and $(x, y) = (0,0)$ at the centre of the cell. Each $G_{m,n}$ has the form of a standing wave whose temporal frequency nearly matches the corresponding hydrodynamic gravity wave mode, whose frequency $f_{m,n}$[18] is independent of current and (in the limit of shallow cells) is given by

$$f_{m,n}^2 = \frac{(\rho_{al} - \rho_c)g}{2\pi \left(\frac{\rho_{al}}{h_{al}} + \frac{\rho_c}{h_c}\right)} \left[\left(\frac{m\pi}{L_x}\right)^2 + \left(\frac{n\pi}{L_y}\right)^2\right]. \tag{2}$$

Here $\rho_{Al}$, $\rho_e$, $h_{Al}$, and $h_e$ are the densities and thicknesses of the Al and electrolyte, respectively. When two modes have nearly identical frequencies and cause surface motion at right angles

(e.g., one with $m = 0$, another with $n = 0$), their resonant coupling can give rise to the MPI[12]. As Eq. 2 shows, if $L_x/L_y = 1$, frequencies of $m = 0$ and $n = 0$ modes match exactly; accordingly, the MPI is predicted to occur even with arbitrarily thick electrolyte layers in square or circular cells. If $L_x \gg L_y$, as in commercial Al reduction cells, $n = 0$ modes have lower frequencies than $m = 0$ modes, and the MPI is avoided if the layer is kept thick, at the expense of reduced energy efficiency and increased carbon emissions.

We simulated the MPI in a TRIMET 180 kA Al reduction cell using the MHD-Valdis software (see Methods). The Al-electrolyte interface developed a central bulge (Fig. 1c) as expected. The cell was stable with a steady, 180 kA current when the electrolyte thickness, quantified by the anode-cathode distance (ACD) was 4.3 cm (Supplementary Fig. 1). With ACD = 4.0 cm, however, the MPI arose (Supplementary Video 1). We quantified MPI magnitude using the root-mean-square (RMS) vertical displacement of the surface from its average shape (see Methods), which has low-frequency dynamics consistent with exponential growth (Fig. 2a). Oscillation is also evident, so we calculated the power spectrum of the displacement at a point, finding its power to be highly concentrated near 0.0263 Hz (Fig. 2b), close to $f_{2,0}$ and $f_{0,1}$, the frequencies of the two modes whose coupling is expected to produce the MPI[12]. To verify the role of those two modes, we performed a least-squares projection of the simulated interface shape onto a basis set of gravity modes $G_{m,n}$, then calculated the root-mean-square (RMS) amplitude $< \alpha_{m,n}^2 >^{1/2}$ of each mode, through the duration of the simulation (Fig. 2c; see Methods). As expected, $G_{2,0}$ and $G_{0,1}$ are far stronger than most other modes, though interestingly, $G_{1,1}$ is also strong. Their amplitudes $\alpha_{m,n}$ grow exponentially over time and oscillate with frequencies near 0.0263 Hz (Figs. 2d-e).

Visualizing the interface shape at four times spaced evenly through a 0.0263-Hz oscillation cycle reveals canonical MPI dynamics: a traveling wave that circulates counter-clockwise when viewed from above (Figs. 2f-i), consistent with the fact that $G_{2,0}$ and $G_{0,1}$ vary with ~90° phase difference. To understand the underlying MPI mechanism, a simplified representation is helpful, so we reconstructed the interface shape after eliminating all modes except $G_{2,0}$ and $G_{0,1}$. The results are consistent with established theory[9,10,12]. Current flows preferentially where the electrolyte is thinner (interface crests), then spreads horizontally after entering the Al layer. Interacting with an ambient vertical magnetic field $B_z$ that points upward at one end of the cell and downward at the other, horizontal current density $j$ produces electromagnetic forces f = j × $B_z$ that drive Al first along the $x$ axis, then along the $y$ axis, forming a closed cycle that amplifies the circulating traveling wave.

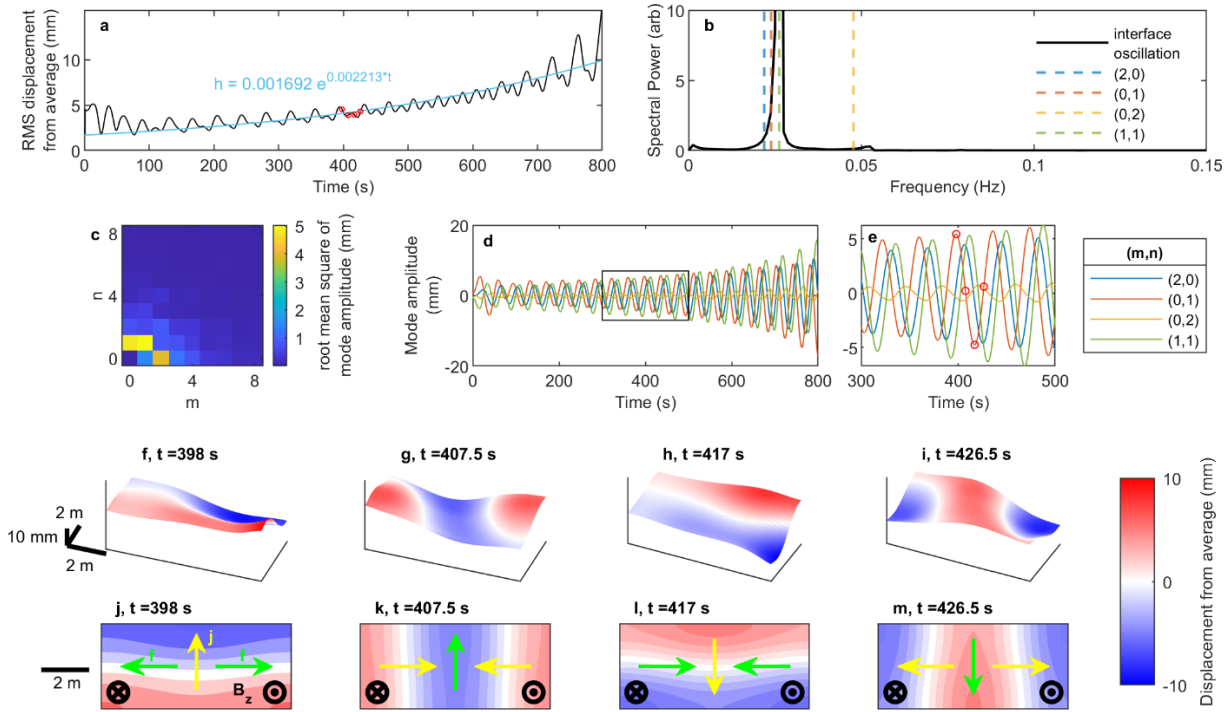

**Figure 2**: The metal pad instability (MPI) in a simulated Al electrolysis cell. **a,** Displacement of the Al-electrolyte interface grows exponentially, as shown by the fitted curve. **b,** The spectral power of the displacement of one point on the interface is dominated by a narrow frequency band, close to the gravitational wave modes $G_{2,0}$ and $G_{0,1}$, expected in the MPI. **c,** $G_{2,0}$, $G_{0,1}$, and $G_{1,1}$ have greater root-mean-square amplitude than any other modes. **d,** $G_{2,0}$, $G_{0,1}$, $G_{1,1}$, and $G_{0,2}$ oscillate with a common frequency and grow over time. **e,** $G_{2,0}$ and $G_{0,1}$ are separated in phase by ~90°, characteristic of a traveling wave as in the MPI. **f-i**, Interface displacements at four times spanning one MPI cycle (red dots in (e)) show a circulating traveling wave. **j-m**, Interface displacements at the same times as in (f-i), estimated using only $G_{2,0}$ and $G_{0,1}$ and viewed from above. The primary directions of the vertical magnetic field $B_z$, horizontal current $j$, and resulting electromagnetic forces $f$ are sketched; their timing and arrangement are right for amplifying the circulating wave.

## Stabilization using an oscillating current

Parametric instabilities can often be decoupled by introducing a new frequency that frustrates the resonance, so we hypothesized that adding an oscillation to the current would prevent the MPI. To test, we ran a new simulation, identical to the one that produced the MPI, except that the steady 180 kA current was supplemented with an oscillating component of magnitude 19.8 kA and frequency 0.045 Hz. Though the RMS interface displacement oscillated, it did not grow exponentially – the MPI was absent (Fig. 3a, Supplementary Video 2). The spectral power of the displacement at a point showed strong peaks at two frequencies, one near $f_{2,0}$ and $f_{0,1}$, as seen in the presence of the MPI, and another near the 0.045 Hz drive frequency (Fig. 3b). Decomposing the interface shape into wave modes, we found that in addition to $G_{2,0}$ and $G_{0,1}$, $G_{0,2}$ and $G_{4,0}$ were also strong (Fig. 3c). Their frequencies $f_{0,2}$ and $f_{4,0}$ nearly match the drive frequency (Fig. 3b), and the temporal variations of their amplitudes $\alpha_{m,n}$ are nearly sinusoidal (Figs. 3d-e), unlike those of $G_{2,0}$ and $G_{0,1}$, which are more complicated. Visualizing the interface shape at four times spaced evenly through a 0.045-Hz oscillation cycle reveals not a traveling wave, as would occur with the MPI, but a standing wave (Figs. 3f-i), consistent with the fact that $G_{0,2}$ and $G_{4,0}$ vary in synchrony and with almost no phase difference (Fig. 3e). Taken together, these facts suggest that a standing gravity wave, driven by the current oscillation, frustrates the circulating wave that comprises the MPI.

For a simplified explanation of the mechanism, we reconstructed the interface shape after eliminating all modes except $G_{0,2}$ and $G_{4,0}$, which account for most of the power and can approximate the actual shape well (Figs. 3j-m). Remembering that current in the Al spreads horizontally from interface crests, we see that the resulting electromagnetic forces tend to drive two vortex-like circulations, one in each half of the cell, whose vorticities alternate over time but are always opposed: one clockwise and one counter-clockwise. Though counter-clockwise flow might tend to facilitate the counter-clockwise MPI circulation (even if not near the cell centre), clockwise flow strongly opposes the MPI and is apparently sufficient, in this case, to prevent it altogether.

These observations suggest a novel strategy for stabilizing Al cells[19]: oscillate the current at a frequency chosen to excite standing wave modes, which frustrate the MPI traveling wave. In addition to the standing wave composed of the $G_{0,2}$ and $G_{4,0}$ modes, we reasoned, other standing wave modes, composed of other $G_{m,n}$ pairs with approximately matched $f_{m,n}$, should also work. In another simulation, we prevented the MPI using a current oscillation of 19.8 kA at 0.069 Hz, exciting $G_{0,2}$ and $G_{6,0}$ (Supplementary Figure 2 and Supplementary Video 3). In another, again using 19.8 kA and 0.045 Hz, we prevented the MPI even with 3.8 cm ACD, demonstrating at least a 12% reduction from the thickness required for stability without oscillation is possible. (Supplementary Figure 3 and Supplementary Video 4). As expected, the average ohmic losses varied approximately linearly with ACD and were reduced 12%. Total average power, which also includes the power necessary for electrolysis, was reduced 4% (Supplementary Figure 4), compared to the case that was stable without current oscillation.

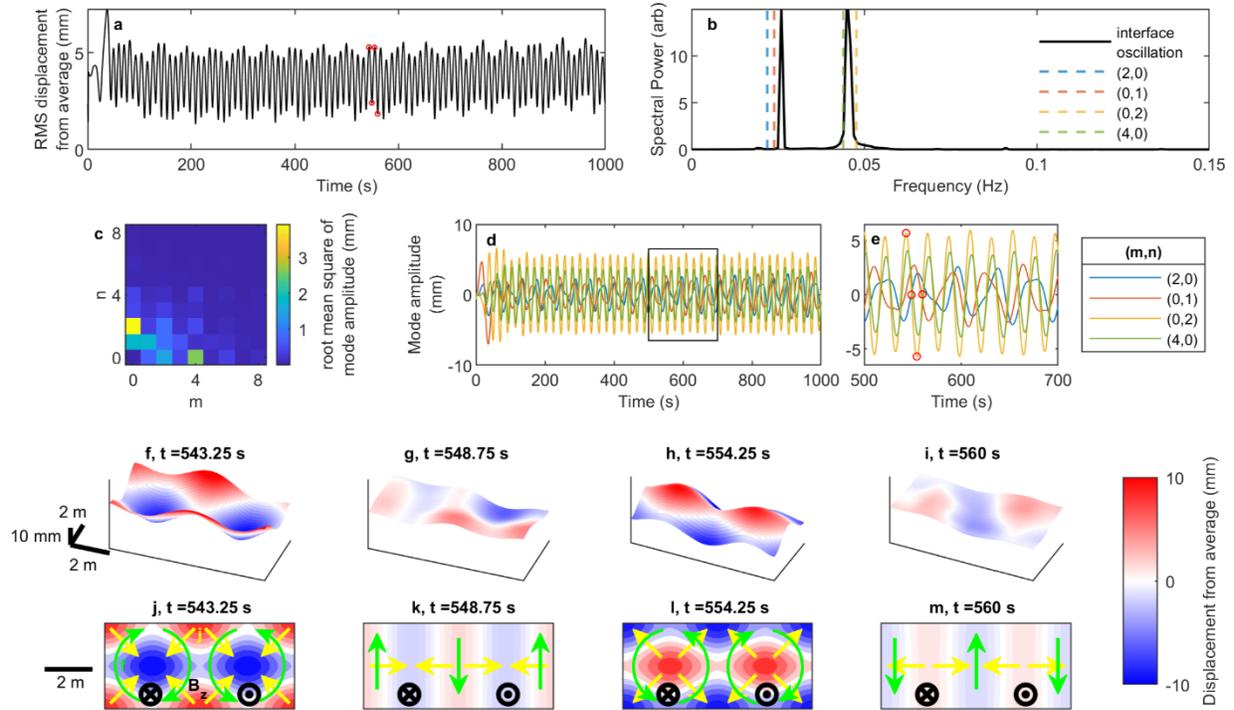

**Figure 3:** An oscillating current component prevents the MPI. **a,** The Al-electrolyte interface oscillates stably. **b,** The spectral power of the displacement of one point on the interface is dominated by one frequency band close to the expected in the MPI frequency, and another close to the drive frequency. **c,** $G_{2,0}$, $G_{0,1}$, $G_{0,2}$, and $G_{4,0}$ have greater root-mean-square amplitude than any other modes. **d-e,** Their amplitudes $\alpha_{m,n}$ oscillate, and $G_{0,2}$ and $G_{4,0}$ are almost aligned in phase, characteristic of a standing wave. **f-i,** Interface displacements at four times spanning one drive cycle (red dots in (e)) show a standing wave. **j-m,** Interface displacements at the same times as in (f-i), estimated using only $G_{0,2}$ and $G_{4,0}$ and viewed from above. The resulting electromagnetic forces $f$ (sketched) often favour clockwise circulation, opposing and frustrating the MPI.

Having prevented the MPI, we wondered if we could also halt it in progress. To find out, we simulated with 4.0 cm ACD, holding the current steady at 180 kA for 100 s before adding a 19.8 kA oscillation at 0.045 Hz (Fig. 4a). As expected, the RMS displacement grew when the current was steady but stopped growing soon after we applied the current oscillation (Figs. 4b and Supplementary Video 5). The interface moved with characteristics of the MPI when the current was steady, but with characteristics of standing waves after we applied the current oscillation (Figs. 4c-j). The MPI was apparently halted.

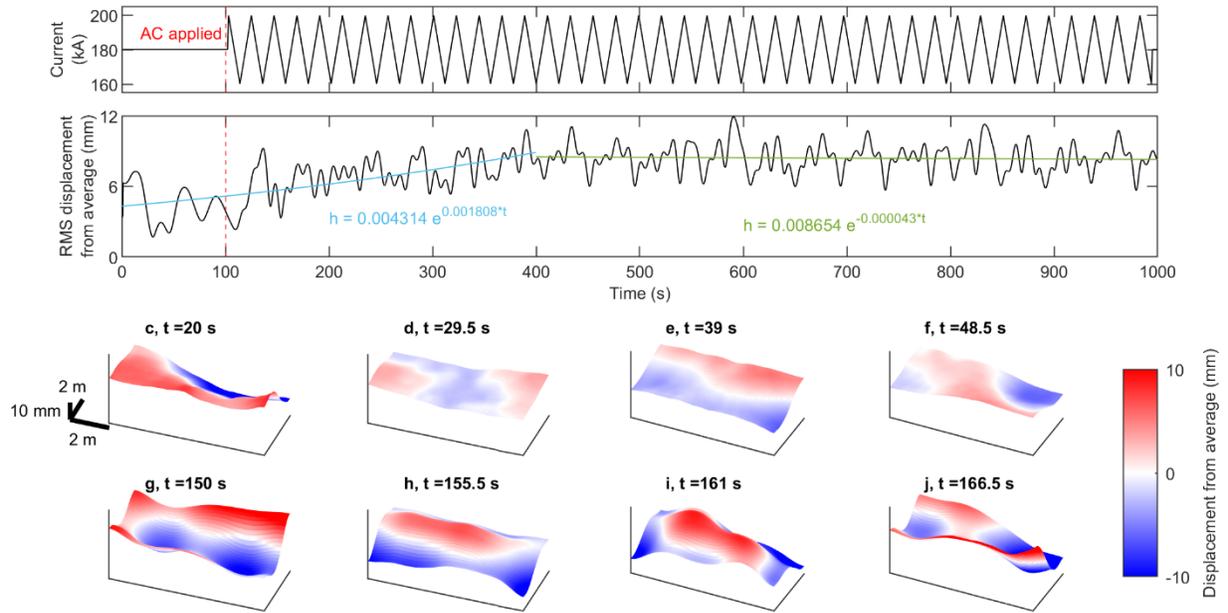

**Figure 4:** An oscillating current component halts the MPI in progress. **a,** The applied current was steady for the first 100 s, before an oscillatory current of 19.8 kA at 0.045 Hz was added. **b,** Interface displacement grew when the current was steady but stopped growing soon after we added current oscillation, indicating stability. **c-f,** Interface displacements at four times spanning one MPI cycle, when current was steady, show a circulating travelling wave. **g-j,** Interface displacements at four spanning one drive cycle, after we added oscillation, show a standing wave.

## Discussion

By applying oscillating currents to excite standing waves, we have both prevented and halted the MPI in simulations of TRIMET 180 kA Al reduction cells. Currently in commercial use by TRIMET Aluminium, SE in Germany, these cells have been simulated in previous studies that validated the model with real-world measurements[20]. Other cell designs differ in, e.g., busbar configuration (affecting $B_z$), size (affecting $f_{m,n}$), and aspect ratio (affecting which modes couple, in the MPI and in standing waves). Still, standing waves composed of low-order modes can be excited in any design by oscillating the current at a frequency determined using Eq. 2, and those standing waves will impose clockwise forces that oppose the MPI much of the time; our novel stabilisation strategy is broadly applicable. For a given design, many standing wave modes are possible, but we speculate that lowest-frequency mode will be strongest for a given oscillation amplitude, because viscous damping is weaker for low-frequency modes. Thus, we hypothesize that low-frequency modes frustrate the MPI with minimal oscillation amplitude; we hope to test that hypothesis in future work. With additional simulations, we found that though the MPI can be prevented with 19.8 kA oscillation, it cannot be prevented with 3.6 kA oscillation. In future work, we hope to prevent the MPI with smaller oscillations. Other designs have been more carefully optimized, allowing stable operation with steady currents and ACD < 3 cm[22]. Still, decreasing ACD in any design could enable producing aluminium with less energy, lower cost, and lower emissions. We hope to test our strategy in an industrial-scale Al cell soon.

Producing the large oscillating currents that we have simulated will require power electronics of substantial cost, but we expect that in nearly all cases, achieving the same efficiency increase by busbar reconfiguration or pot redesign would cost far more. Decreasing ACD by 12% and total power by 4% in all Al reduction cells would reduce the energy intensity of production by 2.1 MJ/kg Al, bringing annual savings of 34 TWh of electricity, perhaps $1B in energy costs, and 13 Mt of $CO_2$e emissions. Future work may yield even bigger decreases. Given that the MPI is likely to occur in liquid metal batteries (a grid-scale energy storage technology)[23-25] and molten oxide electrolysis cells (for electrochemical manufacture of iron and other metals)[26,27], it may be useful to apply oscillating currents in those systems as well.

**Acknowledgments**

The authors are grateful for fruitful discussions with Riccardo Betti, Curtis Broadbent, Gerrit M. Horstmann, and Jonathan S. Cheng. This work was supported by the National Science Foundation (CBET-1552182) and by a University of Rochester, URVentures TAG award.

# Methods

**Simulation software**

To study the stability of a typical aluminium reduction cell under the influence of an oscillating AC current, we used the simulation package MHD-Valdis, a tool widely used in the industry to design stable aluminium reduction cells. MHD-Valdis has been described in detail[28]. It dynamically couples the transient turbulent motion of each fluid layer and the interface shape to the transient magnetic field and electric currents in the cell[28]. The model includes essential commercial cell features such as the electrolyte channels[29], electric current distribution in the bus bars, and the magnetic field generated by the ferromagnetic cell elements[28], and has been validated against a benchmark model[30] and against measurements from the TRIMET 180 kA commercial potline[20] and other commercial cells[31,32]. A model of the same TRIMET cell is used in our stabilization study. MHD-Valdis uses the shallow-layer approximation in the Al and electrolyte layers, and utilizes a k-ω turbulence model. The electric boundary conditions used are zero normal current at the cell's side walls, a specified current density distribution at the anode and cathode blocks that changes in time with the interface shape, continuity of electric potential across the Al-electrolyte interface, and continuity of the normal current component across the Al-electrolyte interface. Additionally, the hydrodynamic boundary conditions used are continuity of pressure across the Al- electrolyte interface, zero normal velocity at the side walls, and zero horizontal circulation velocity $\boldsymbol{u}_h$ at the side walls where $\boldsymbol{u}_h$ comes from decomposing the velocity field:

$$\boldsymbol{u}(x,y,z,t) = \boldsymbol{u}_h(x,y,t) + \epsilon\, u_w(x,y,z,t) + O(\epsilon^2) \qquad (3)$$

Where $\epsilon$ is a non-dimensional parameter that is the ratio of the wave amplitude to the depth of the fluid layers (assumed very small), and $u_w$ is the wave-related velocity.

**Simulation protocol**

The interface shape is initialized with the $G_{1,0}$ gravity mode as a perturbation with an amplitude of $\alpha_{1,0} = 5$ mm, similar to previous work[20], and a steady and/or oscillating current with a specified frequency and amplitude. We use a 0.25 s time step and a numerical grid having $N_x \times N_y = 87 \times 31$ elements, so the grid size is $11.90 \times 9.21$ cm. Simulation is halted if the interface touches the anode, or after a set time has elapsed. A series of simulations were done with purely steady current but at varying ACD to find its critical value for stability. At the critical ACD, the motion of a point on the interface has an amplitude that neither grows nor decays in time. We then ran a simulation at ~5% below the critical ACD and confirmed that the cell was unstable by observing the interface amplitude growing exponentially in time. When applying oscillating currents, we used a triangular shape (Fig. 4a) because it may be more economical to produce in real Al cells and because it eliminated the need for shorter time steps in simulations.

**Projecting onto wave modes**

To express the interface displacement in terms of gravity wave modes, we used linear least-squares projection onto an orthonormal basis set formed from the $G_{m,n}$ modes defined in Eq. 1. In brief, linear least-squares projection is performed by determining the amplitudes $\alpha_{m,n}$ that minimize the squared error between the measured surface displacement and the surface displacement estimated using Eq. 1. We retained the 128 lowest-order modes ($0 \leq m \leq 16$ and $0 \leq n \leq 8$). Spectral power in modes with $m > 6$ or $n > 6$ was always negligible. The $\alpha_{m,n}$ vary over time; one way to quantify their typical overall magnitude is with the root-mean-square amplitude $<\alpha_{m,n}^2>^{1/2}$, where brackets signify averaging over time.

**Quantifying surface displacement**

We calculated the root-mean-square (RMS) displacement of the surface from its mean as

$\left(\frac{1}{N_x N_y}\sum_{i=1}^{N_x}\sum_{j=1}^{N_y}(z(i,j)-\bar{z}(i,j))^2\right)^{1/2}$, where the simulation grid has $N_x \times N_y$ elements, $z(i,j)$ is the surface height of an element, and $\bar{z}(i,j)$ is its time-averaged height. When considering frequencies present in the surface displacement, we did not use the RMS, but rather used a single point in one corner of the cell, without squaring its displacement, since the square of a function generally contains different frequencies than the original function. We calculated power spectra using the Fast Fourier Transform (fft).

contributions

**Author contributions**

M.D. conducted the simulations. P.D.F. guided experimental design choices for the simulation campaign. I.M. and D.H.K. wrote the manuscript and performed data analysis, including representation in term of wave modes. D.H.K. conceived the study.

**Competing interests**

The authors declare no competing interests.

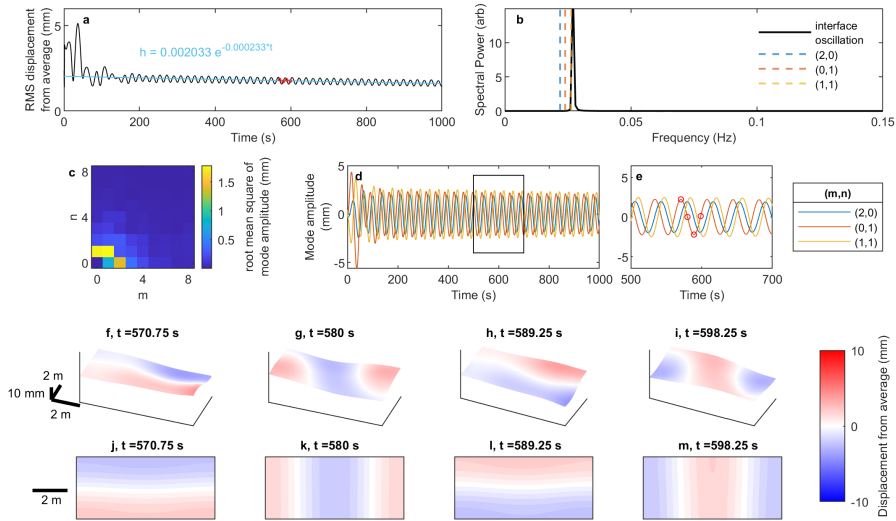

**Supplementary Figure 1:** Stable cell with steady 180 kA current and 4.3 cm ACD . **a,** The Al-electrolyte interface oscillates and decays in time, as shown by fitted curve. **b,** The spectral power of the displacement of one point on the interface is dominated by one frequency band close to the gravitational wave modes $G_{2,0}$ and $G_{0,1}$. **c,** $G_{2,0}$, $G_{1,1}$ and $G_{0,1}$ have greater root-mean-square amplitude than any other modes. **d-e,** Their amplitudes $\alpha_{m,n}$ oscillate **f-i,** Interface displacements at four times spanning one drive cycle (red dots in (e)) show a circulating wave. **j-m,** Interface displacements at the same times as in (f-i), estimated using only $G_{0,2}$ and $G_{1,0}$ and viewed from above. The flow perturbation applied at the beginning of the simulation excites the same wave modes that arise in the unstable case with 4.0 cm ACD (Fig. 2), but instead of growing stronger over time, they decay, demonstrating that the cell is stable.

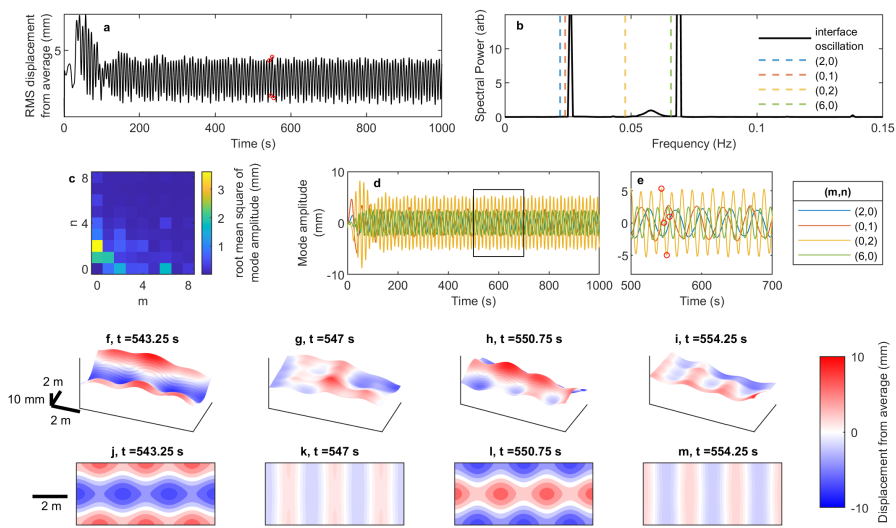

**Supplementary Figure 2:** An oscillating current component prevents the MPI. **a**, The Al-electrolyte interface oscillates stably. **b**, The spectral power of the displacement of one point on the interface is dominated by one frequency band close to the expected in the MPI frequency, and another close to the drive frequency. **c,** $G_{2,0}$, $G_{0,1}$, $G_{0,2}$, and $G_{6,0}$ have greater root-mean-square amplitude than any other modes. **d-e,** Their amplitudes $\alpha_{m,n}$ oscillate **f-i,** Interface displacements at four times spanning one drive cycle (red dots in (e)) show a standing wave. **j-m,** Interface displacements at the same times as in (f-i), estimated using only $G_{0,2}$ and $G_{6,0}$ and viewed from above.

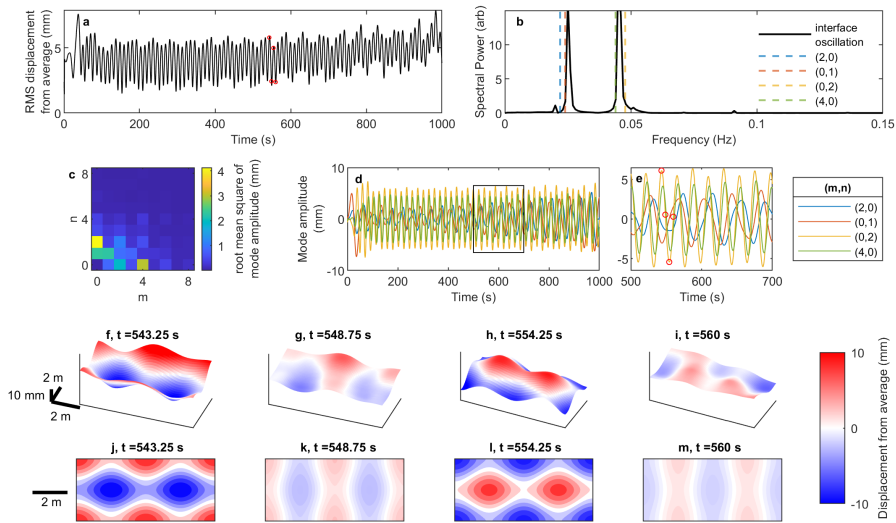

**Supplementary Figure 3:** An oscillating current component prevents the MPI. **a,** The Al-electrolyte interface oscillates stably. **b,** The spectral power of the displacement of one point on the interface is dominated by one frequency band close to the expected in the MPI frequency, and another close to the drive frequency. **c,** $G_{2,0}$, $G_{0,1}$, $G_{0,2}$, and $G_{4,0}$ have greater root-mean-square amplitude than any other modes. **d-e,** Their amplitudes $\alpha_{m,n}$ oscillate **f-i,** Interface displacements at four times spanning one drive cycle (red dots in (e)) show a standing wave. **j-m,** Interface displacements at the same times as in (f-i), estimated using only $G_{0,2}$ and $G_{4,0}$ and viewed from above.

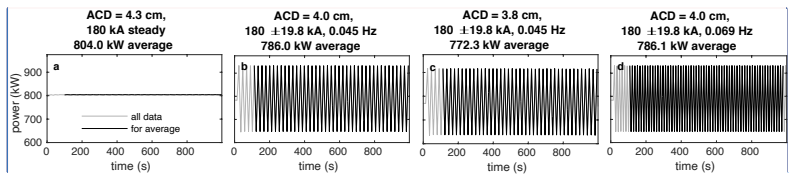

**Supplementary Figure 4:** Average and instantaneous power in simulations with varying ACD and current oscillation. **a,** Applying a steady, 180 kA current, the minimum ACD for which the cell is stable is 4.3 cm, and under those conditions, the average power is 804.0 kW. **b-d,** Applying an oscillating current component in addition to the 180 kA steady component allows stable operation at smaller ACD, and correspondingly lower power. Among the cases considered, the minimum average power is 772.3 kW with a 3.8-cm ACD and a 19.8 kA oscillation at 0.045 Hz, representing a 4% power reduction. To calculate average power, we average over an integer number of oscillations, starting after the first 100 s to ensure the simulation has stabilized.

| Electrical conductivity of liquid aluminium | 4000000 S/m |
| Electrical conductivity of liquid electrolyte | 233 S/m |
| Electrical conductivity of anode carbon | 20450 S/m |
| Electrical conductivity of cathode carbon | 74074 S/m |
| Density of liquid aluminium | 2300 kg/m^3 |
| Density of liquid electrolyte | 2075 kg/m^3 |

**Table 1:** Material properties used in the simulations.